\documentclass[11pt]{article}
\usepackage{graphicx}

\newcommand{\BABARPubYear}    {01}

\newcommand{\BABARConfNumber} {92}
\newcommand{\SLACPubNumber} {9048}

\input pubboard/babarsym

\begin{document}
{\pagestyle{empty}

\begin{flushright}
\babar-CONF-\BABARPubYear/\BABARConfNumber \\
SLAC-PUB-\SLACPubNumber \\
November, 2001 \\
\end{flushright}

\par\vskip 3cm

\begin{center}
\Large \bf \boldmath
Recent Results of \babar\ 
\end{center}
\bigskip

\begin{center}
D.~Bernard\footnote{Email: denis.bernard@in2p3.fr}, \\
Ecole Polytechnique, Palaiseau, France\\
For the {\bf \boldmath \babar\ Collaboration}\\
\end{center}
\bigskip 

\begin{center}
Presented at ``{\bf New Trends in High-Energy Physics}'' \\
         Yalta, Crimea, Ukraine \\
            September 22 -- 29, 2001
\end{center}
\bigskip  \bigskip  \bigskip  \bigskip

\begin{center}
\large \bf Abstract
\end{center}
The  \babar\ detector at SLAC's  \pep2\ storage ring 
has collected data equivalent to about 30.4~\invfb through June 2001.
Results on \CP violation, and in particular searches for direct \CP violation,
and measurement of rare \B decays are presented.

\vfill
\begin{center}
{\em Stanford Linear Accelerator Center, Stanford University, 
Stanford, CA 94309} \\ \vspace{0.1cm}\hrule\vspace{0.1cm}
Work supported in part by Department of Energy contract DE-AC03-76SF00515.
\end{center}
\bigskip \bigskip \bigskip \bigskip

}

\newpage

\setcounter{footnote}{0}

\section{Introduction}
\label{sec:Introduction}

We present a sample of recent results from \babar\,
including the observation of \CP violation with the measurement of \stwob
and of rare decay modes.

The \babar\ experiment has been running at the \pep2\ asymmetric \epem
collider since 1999.
Its main goal is a high statistics study of \B decays, including a study of \CP
violation in the \B sector.
The center-of-mass energy is tuned to the \FourS just above the \BB threshold.
The asymmetry of the energies of the two beams provides a longitudinal boost so that
the average \B flight length is $\approx$ 250 \mum and can be measured.
The high nominal luminosity of $3\times 10^{33}$ \cms allows the study
of many rare \B decay channels.


The \babar\ detector is described elsewhere~\cite{ref:babar}.
Charged particle track parameters are obtained from measurements in a
5-layer double-sided silicon vertex tracker and a 40-layer drift
chamber located in a 1.5-T magnetic field; both devices provide \dedx\
information. Additional charged particle identification (PID)
information is obtained from a detector of internally reflected
Cherenkov light (DIRC) consisting of quartz bars that carry the light
to a volume filled with water, and equipped with 10752 photomultiplier
tubes. Electromagnetic showers are measured in a calorimeter (EMC)
consisting of 6580 CsI(Tl) crystals. 
An instrumented flux return (IFR), containing multiple layers of
resistive plate chambers, provides $\mu$ identification.

Most results presented here are obtained with a data sample of 30.4
\invfb collected through June 2001, for a total amount of  about $32\times 10^{6} \BB$ pairs.

\section{Measurement of \boldmath \stwob}
\label{sec:sin2beta}

The primary goal of the experiment is the observation of \CP violation in the \B sector.
In the standard model (SM), \CP violation occurs via a complex term in the
Cabibbo-Kobayashi-Maskawa (CKM) matrix.
The phase of such a complex term can eventually be measured from the
interference of two amplitudes contributing to the same final state.
In the Wolfenstein parametrization\cite{ref:wolf} of $V_{CKM}$,
all the matrix elements are real, except $V_{td}$ and $V_{ub}$.
Most of the envisaged strategies to observe \CP violation therefore
use the interference between an amplitude containing one of these target
complex CKM
elements and a real amplitude, to a \CP eigenstate.

In the case of a $b \to c \bar{c} s$ decay like $\B \to \jpsi \KS$,
the \B can either decay directly via a real amplitude $\propto
V_{cb}^*V_{cs}$, or decay after a \Bz $\rightarrow$ \Bzb oscillation
that proceeds through a box diagram with an amplitude
$(V^*_{tb}V_{td})^2$ with a phase $-2\beta$.
The measurement is theoretically clean, as the main higher-order
diagrams have the same weak phase as the first-order ones.
It is also relatively background free due to the
presence of a \jpsi in the final state, and benefits from a branching
ratio \BR $\approx 10^{-3}$, a rather high value in the land of \B
decays.

The resulting evolution function $g_{\pm}(t)$, for a decay to a final
state $f$, involves a complex quantity $\lambda_{f}$.
If the (so called direct) \CP violation in the decay itself can be
neglected -- as it is the case here, within 1~\%, in the standard
model -- we have $|\lambda_{f}|=1$, and we obtain:
\begin{eqnarray*}
g_{\pm}(t)
= \frac{e^{- t / \tau_{\B}}}{\tau_{\B}} \times 
\left[1 \pm {\mathcal Im} \lambda_{f} \sin (\deltamd t) \right],
\end{eqnarray*}

\noindent where $t$ is the \B proper decay time, and the $+$ or $-$
 corresponds to an intial \Bzb\ or \Bz\ meson.
In the present case of a $b \to c \bar{c} s$ decay, $\lambda_{f}$ is
$\eta_{f} e^{-2i\beta} $, where $\eta_{f}$ is the \CP eigenvalue of
$f$, and $g$ simplifies to:
\begin{equation}
g_{\pm}(t)
= \frac{e^{- t / \tau_{\B}}}{\tau_{\B}} \times 
\left[1 \mp \eta_{f} \stwob \sin (\deltamd t) \right],
\label{eq:eveq}
\end{equation}

\noindent
allowing the measurement of \stwob.
The ``initial'' state of the \CP \B can be known thanks to the EPR
paradox: the \FourS resonance having spin-parity $1^{--}$,  decays to a coherent
\BB pair in an antisymmetric state, so that when the first \B decays, say
to a \Bz, the other one is in the opposite state, a \Bzb.
The oscillation time is then the difference $\Delta t$ of the time
of flight of the two \B's, measured from the difference $\Delta z$ of their flight
lengths.
The vertexing of the \CP \B is performed with an excellent precision,
\sz $\approx$ 60 \mum, thanks to the presence
of the hard leptons from the \jpsi.
The other \B is vertexed inclusively, with a $\chi^2$ cut that limits 
the systematics due to cascade charmed decays; the contribution 
of the other \B dominates the resolution on $\Delta z$, which is
of the order of 180 \mum independent of the \CP channel used.

The flavor determination of the other \B, called tagging, is also
performed inclusively: events are classified in four mutually-exclusive
categories, using respectively the charge of the fastest lepton, the
total charge of identified kaons, the output of neural networks that
 use the information carried by non-identified leptons and
kaons, and by soft pions from \Dstar's, or are not tagged \cite{ref:prlfeb2001}.
The effective tagging efficiency $Q_{i}$ of each category $i$ is
defined as the product of its efficiency $\epsilon_{i}$ and of the
square of its dilution $(1-2w_{i})$.
The total effective tagging efficiency $Q = \sum_{i=1}^{4}
\epsilon_{i} (1-2w_{i})^{2}$ is measured on a large sample of 
 exclusively reconstructed \B decays of specific flavor
($\B \to D^{(*)-} h^{+}, h=\pi,\rho,a_{1}$, and $\B \to \jpsi
\Kstarz (\Kp \pim)$), and is equal to $(26.1 \pm 1.2)~\%$.

With a real detector, \stwob is multiplied by a dilution factor
 $(1-2w)$ in eq. \ref{eq:eveq}, and 
$g_{\pm}(\Delta t)$ is convoluted with a $\Delta t$ resolution function.
The value of \stwob is obtained from an unbinned maximum likelihood
fit to the $\Delta t$ distribution of a combined event sample
consisting of the \CP sample and of the flavor sample.
(The number of tagged events, the purity, and the \CP effective value
of the channels used are given in table \ref{table:chans}).
In this way, the mistagging probabilites $w_{i}$ 
and the parameters of the resolution function are determined from the data,
and their correlation with \stwob is taken into account.

\begin{figure}[!htb]
\begin{center}
\includegraphics[width=0.6\linewidth]{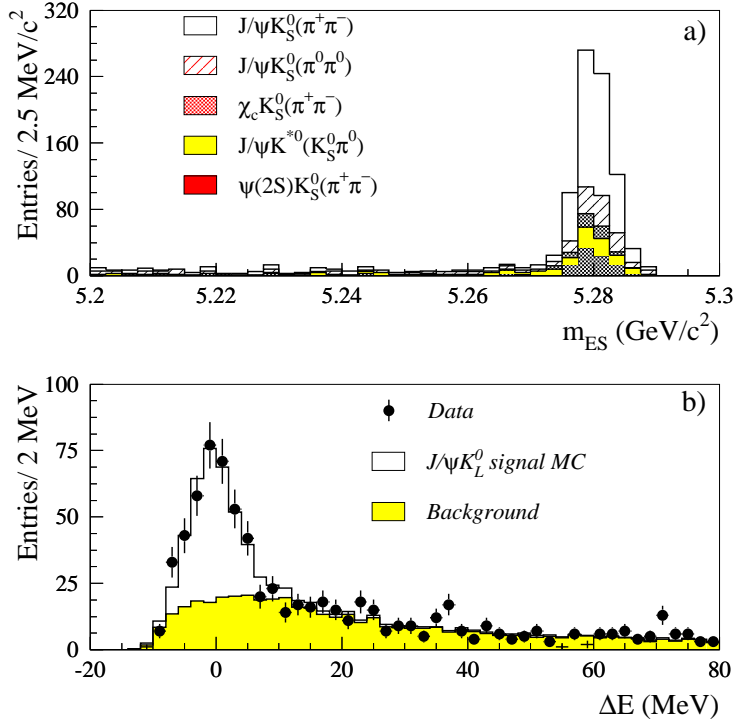}
\caption{
 a) Distribution of \mes\ 
for \B candidates 
having a \KS 
in the final state;
b) distribution of $\Delta E$ for $\jpsi\KL$ candidates.
\label{fig:mesde}}
\end{center}
\end{figure}

Background events are taken into account by a separate {\em pdf} that
enters the likelihood. 
When the \CP \B has no \KL in the final state, the probability for an
event to be a signal event is computed from the value of the energy
substitued mass $\mes = (s/4 - {\vec{p}_B^{*2}})^{1/2}$ where $s$ is the
total energy and $\vec{p}_B^{*}$ the measured momentum of the
\B candidate, in the \FourS rest frame (fig. \ref{fig:mesde}a).
For a \CP \B with a \KL in the final state, only the direction of the
\KL is measured in the detector, and a kinematic fit of \mes to the nominal \B mass
is performed. 
The signal probability of the event is computed from the difference
$\Delta E = E^*_B - E^*_{beam}$ between the energy of the \B candidate
and the beam energy, in the \FourS\ rest frame (fig. \ref{fig:mesde}b).

\begin{table}[!htb]
\caption{
Number of tagged events, signal purity and \CP content of 
the \CP and flavor samples.
\label{table:chans}}
\begin{center}
\begin{tabular}{c|ccccc}
 & Channel & $N_{tag}$ & Purity (\%)
& $ \langle\eta_{f} \rangle$ \\
\hline \hline
CP sample &
 \jpsi\KS (\pip\pim) & 316 & 98. & $-1$\\
& \jpsi\KS (\piz\piz) & 64 & 94. & $-1$\\
& \psitwos \KS (\pip\pim) & 67 & 98. & $-1$\\
& \chicone \KS (\pip\pim) & 33 & 97. & $-1$ \\
& \jpsi \Kstarz (\KS\piz) & 50 & 74. & $0.65 \pm 0.07$ \\
& \jpsi\KL & 273 & 51. & +1 \\
\hline \hline
 non \CP sample & Flavor & 7591 & 86 & \\
\hline \hline
\end{tabular} 
\end{center}
\end{table}

\begin{figure}[!htb]
\begin{center}
\includegraphics[width=0.4\linewidth]{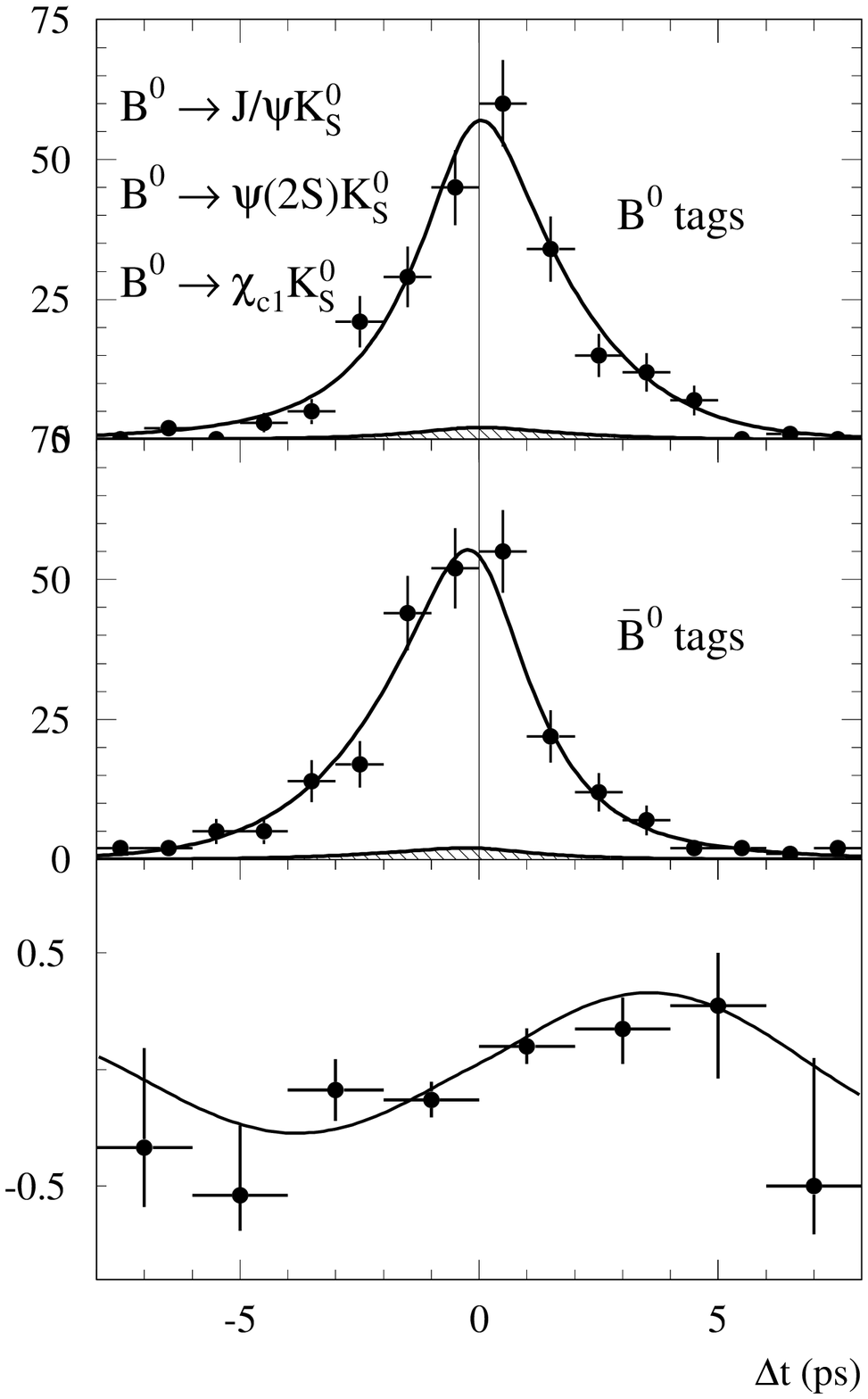}
\includegraphics[width=0.4\linewidth]{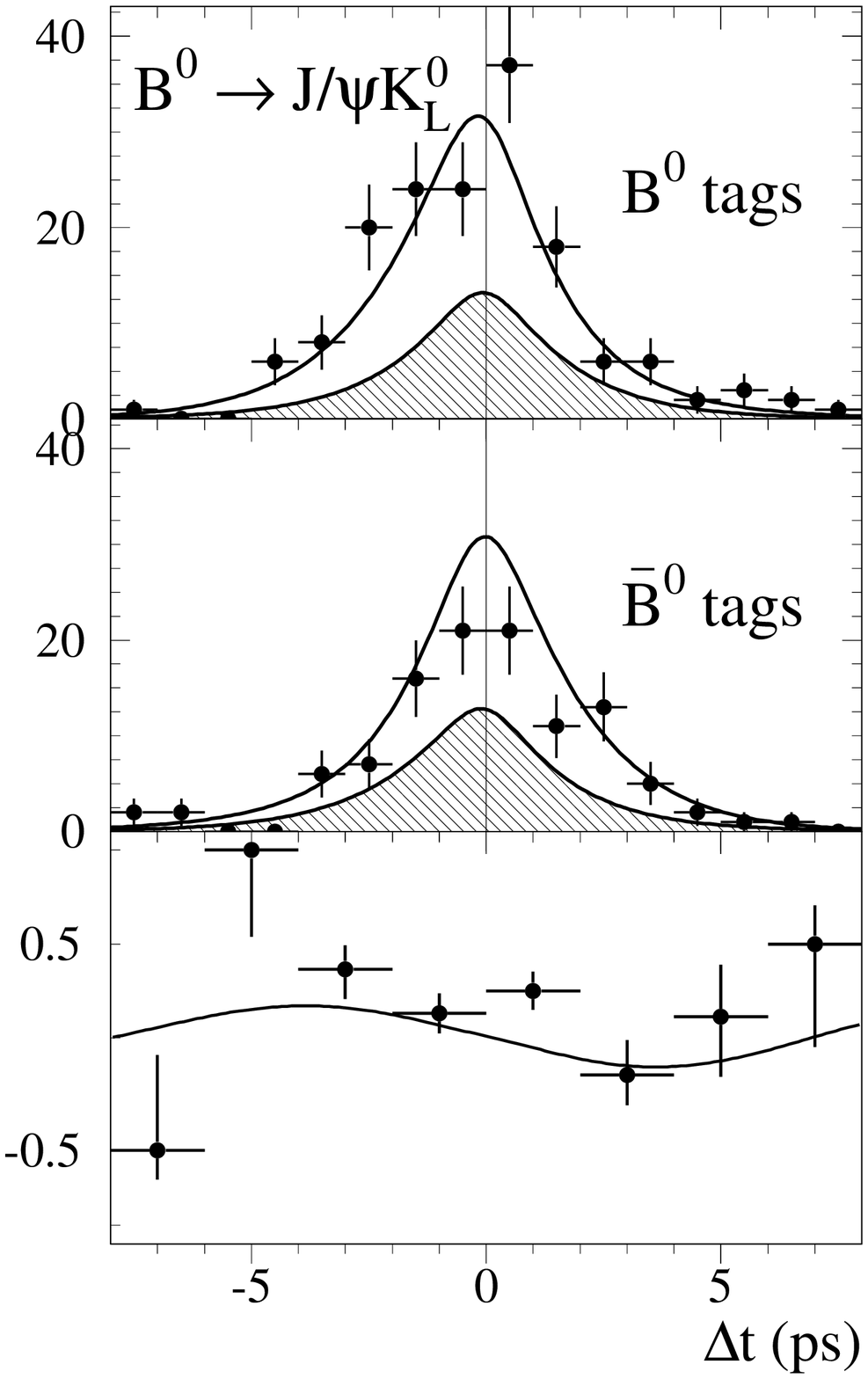}
\caption{Left:
Time distribution of $\eta_f=-1$ candidates with a \Bz tag
$N_{\Bz}$ and with a \Bzb tag $N_{\Bzb}$, and the asymmetry
$(N_{\Bz}-N_{\Bzb})/(N_{\Bz}+N_{\Bzb})$. The solid curves represent
the result of the combined fit to all selected \CP events; the shaded
regions represent the background contributions. Right:
Corresponding information for the $\eta_f=+1$ mode $(\jpsi\KL)$. 
\label{fig:dt}}
\end{center}
\end{figure}

The slight asymmetry of the time distributions (fig.~\ref{fig:dt}) is
barely visible, but the effect is clear in the plot of the asymmetry
itself 
${\mathcal A}_{\CP}(\Delta t) =
 -(1-2w)\eta_{f} \sin\! 2 \beta \sin (\deltamd \Delta t)$.
We obtain $\stwob = 0.59 \pm 0.14 (stat) \pm 0.05 (syst)$. If there
were no \CP violation, i.e. $\beta = 0$, the probability of such a
measurement would be $3\times 10^{-5}$.
Therefore, our result for \stwob represents a 4.1 standard deviation
observation of \CP violation \cite{ref:prlsept2001} in the neutral \B meson system.

\begin{figure}[!htb]
\begin{center}
\includegraphics[width=0.4\linewidth]{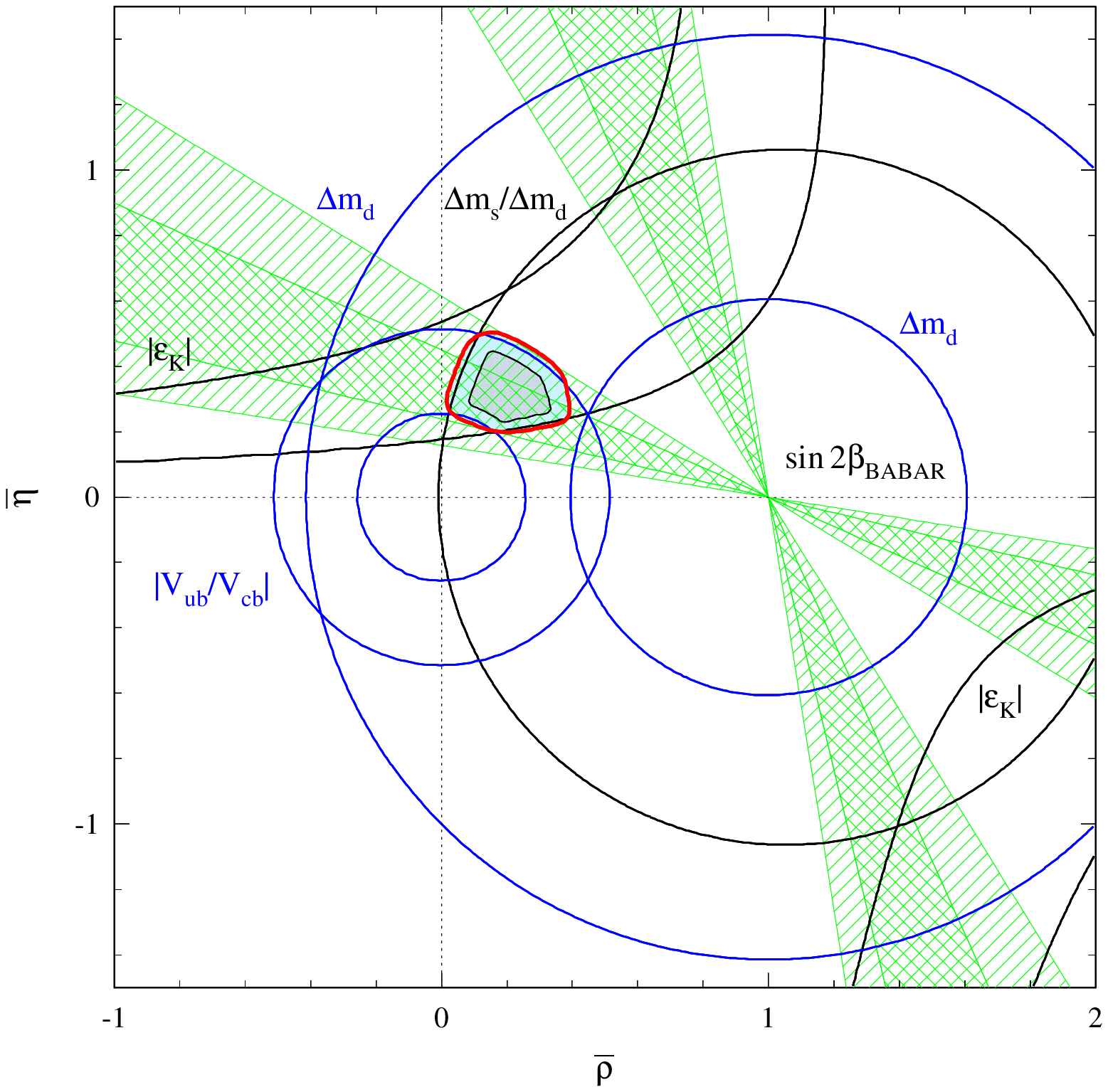}
\includegraphics[width=0.4\linewidth]{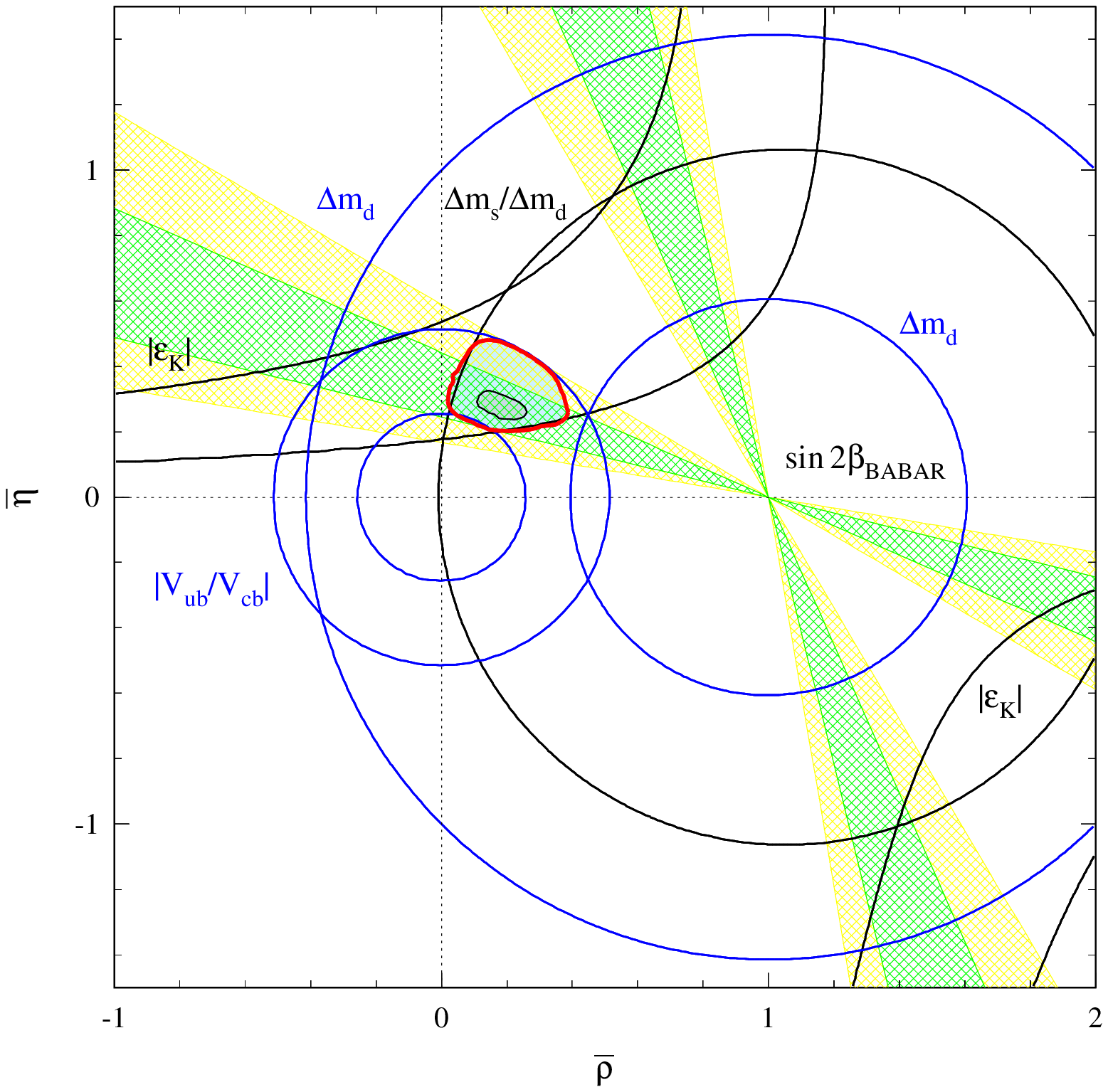}
\caption{Allowed region 
(light (dark) grey: two (one) standard deviation)
in the CKM parameters $\rho, \eta$ plane.
Left: the present \stwob measurement is overlaid on preceding fit.
Right: the present \stwob measurement is included in the fit.
The fit method is described in ref. \protect\cite{ref:andreas}.
\label{fig:ckm}}
\end{center}
\end{figure}

The impact of the present measurement on
the knowledge of the CKM matrix is shown in fig. \ref{fig:ckm}.
The set of measurements is clearly consistent within the framework of
the standard model, while the present measurement significantly
decreases the size of the $2\sigma$ allowed region.

We have also searched for direct \CP violation in the decay. Releasing
the constraint
\footnote{but still assuming that there is no \CP violation in mixing.}
$|\lambda_{f}|=1$, the evolution equation becomes:
\begin{eqnarray*}
 g_{\pm}(\Delta t) =
\frac{e^{- |\Delta t |/ \tau_{\B}}}{2 \tau_{\B}
(1+|\lambda_{f}^2|)} \times 
\left[ \frac{1+ \vert \lambda_{f} \vert^2}{2}
 \pm \left( - \frac{1}{2}(1 - \vert \lambda_{f} \vert^2)
\cos (\deltamd \Delta t) + {\mathcal Im} \lambda_{f} \sin(\deltamd \Delta t) \right) \right] 
\end{eqnarray*}

Fitting for $|\lambda_{f} |, {\mathcal Im}\lambda_{f}/|\lambda_{f}|$
with the $\eta_{f}=-1$ sample, we obtain\cite{ref:prlsept2001}
$ |\lambda_{f} | = 0.93 \pm 0.09 \pm 0.03 $,
showing the absence of direct \CP violation at the 10\% level.

\section{Charmless two body \B decays}

The $V_{ub}$ term that appears in the amplitude of the decay $b \to u$
carries a $\gamma$ phase: this makes the measurement of \stwoa in
$B\to \pip\pim$ decay-mixing interference attractive, using a time
distribution analysis similar to that for \stwob.
However, in this case the decay has a strong CKM suppression ($V_{ub}
\propto \lambda^{3}$), so that the higher-order diagrams, so called penguins,
contribute at the same level as the tree diagram. This leads to a
complication in the measurement of \stwoa, but also makes the process sensitive
to the contribution of new heavy objects (Higgs, SUSY) in the penguin diagram,
which would result in direct \CP violation via tree-penguin interference in
$\B\to \Kp\pim$ decays.


On the experimental side, the difficulty is background rejection. The
decaying \B's are almost at rest in the \FourS rest
frame, and their 2-body decay products are approximately back-to-back
in that frame, while continuum ($\epem \to \qqbar$) events have
a two-jet topology that provides a lot of back-to-back track pairs.
\BB decay products are roughly isotropically distributed in the \FourS frame
and we use this behaviour to separate signal from noise.

Preselection cuts on global event shape parameters (Fox-Wolfram
moment, sphericity), on the angle between the sphericity axes of the \B
and that of the rest of event, and on a Fisher discriminant ${\cal F}$
that optimizes the use of the energy flow with respect to the decay axis
\cite{ref:charmlessAsym}.
Kaon/pion separation is achieved using the 
Cherenkov angle $\theta_c$ measured in the DIRC and the value of 
$\Delta E$ (computed with the pion hypothesis for both tracks).

\begin{figure}[!htb]
\begin{center}
\includegraphics[width=0.4\linewidth]{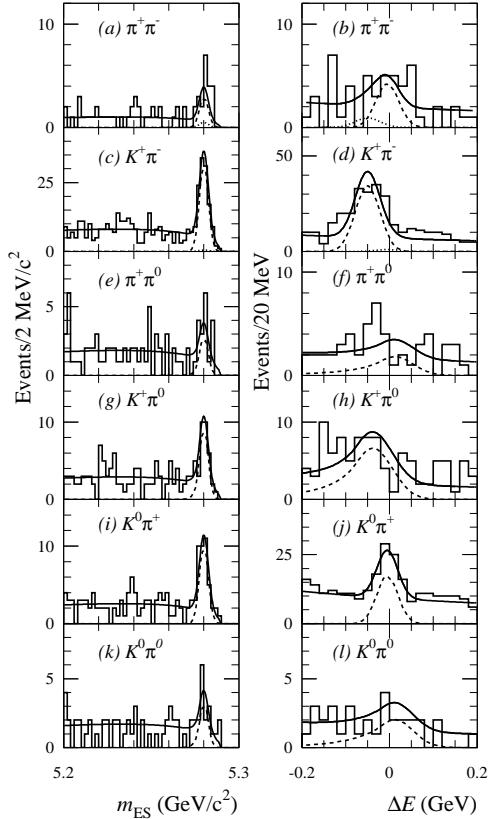}
\caption{The \mes\ and $\Delta E$ distributions of charmless two body decays,
after a likelihood ratio cut. The solid curves represent the fit
predictions for both signal and background; the dashed curve
represents the given signal mode only and the dotted curve represents
other modes of the same topology (20.7~\invfb).
\label{fig:charmless}}
\end{center}
\end{figure}

The branching fractions are obtained by an unbinned maximum likelihood
fit with the 4 variables: \mes, $\Delta E$, ${\cal F}$, $\theta_c$
and are given in table \ref{table:charmless}.
\begin{table}[!htb]
\caption{Detection efficiencies
($\varepsilon$), fitted signal yields ($N_S$), statistical
significances ($S$), branching fractions (\BR), and charge asymmetries
of the charmless two body decays
(20.7~\invfb).
\label{table:charmless}}
\begin{center}
\begin{tabular}{lcccccc} 
\hline\hline
Mode  & $\varepsilon$ (\%) & $N_S$ & $S$ ($\sigma$) & ${\cal B}/10^{-6}$ &
$A_{CP}$ \\
\hline
$\pip\pim$ &  $45$  & $41\pm 10\pm 7$ & $4.7$  & $4.1\pm 1.0\pm 0.7$\\
$\Kp\pim$ &  $45$ & $169\pm 17\pm 13$ &$15.8$ & $16.7\pm 1.6\pm 1.3$ &   $-0.19\pm 0.10\pm 0.03$ \\
$\Kp \Km$  & $43$ & $8.2^{+7.8}_{-6.4}\pm 3.5$ & 1.3 &  $<2.5$ ($90\%$ C.L.)   \\
$\pip\piz$  & $32$ & $37\pm 14\pm 6$ & $3.4$  & $<9.6$ ($90\%$ C.L.) \\
$\Kp\piz$    & $31$ & $75\pm 14\pm 7$ & $8.0$  & $10.8^{+2.1}_{-1.9}\pm 1.0$ & $0.00\pm 0.18\pm 0.04$  \\
$\Kz\pip$   & $14$ & $59^{+11}_{-10}\pm 6$ &$9.8$  & $18.2^{+3.3}_{-3.0}\pm 2.0$   & $-0.21\pm 0.18\pm 0.03$   \\
$\Kzb\Kp$    & $14$ & $-4.1^{+4.5}_{-3.8}\pm 2.3$  &$-$    & $<2.4$ ($90\%$ C.L.) \\
$\Kz\piz$  & $10$ & $17.9^{+6.8}_{-5.8}\pm 1.9$ &$4.5$  & $8.2^{+3.1}_{-2.7}\pm 1.2$ \\
\hline\hline
\end{tabular}
\end{center}
\end{table}
The corresponding \mes and $\Delta E$ spectra, after appropriate cut
on the likelihood ratio to enhance the signal fraction, are given in
fig. \ref{fig:charmless}.
Self-tagging modes are used to search for direct \CP violation in the charge asymmetry
$A_{CP}= (\BR(\Bbar\to\bar{f}) -\BR(\B\to f))/(\BR(\Bbar\to\bar{f}) +\BR(\B\to f) )$.


A vigorous effort is under way to understand the shift in \stwoa\ 
due to the presence of the penguins\cite{ref:stwoa:th}.
The quantity that is actually measured is denoted $\stwoa_{eff}$.

An unbinned maximum likelihood fit similar to that used for the branching fractions
is used to measure  $\stwoa_{eff}$:
 the variable $\Delta t$ is added in the likelihood.
The vertexing and tagging are similar to that used for \stwob.
There is here a (strong) probability of direct \CP violation so that the
evolution equation used is:
\begin{eqnarray*}
 g_{\pm}(t) =
\frac{e^{- |\Delta t |/ \tau_{\B}}}{4 \tau_{\B}}
 \times 
\left[ 1 \pm \left( -C_{f}\cos (\deltamd t) + S_{f} \sin(\deltamd t) \right) \right] 
\end{eqnarray*}

Fitting for the coefficients 
$\displaystyle
C_{f}= {(1 - \vert \lambda_{f} \vert^2)}/{(1 + \vert \lambda_{f} \vert^2)}
$
and
$\displaystyle
S_{f}= {2{\mathcal Im} \lambda_{f}}/{(1 + \vert \lambda_{f} \vert^2)}
$
we get \cite{ref:stwoa}, on 30.4 \invfb,
$S_{f} = 0.03^{+0.53}_{-0.56} \pm 0.11$
and
$C_{f} = -0.25^{+0.45}_{-0.47} \pm 0.14$ (preliminary),
where $S_{f}$ would be equal to  $\stwoa_{eff}$
if $|\lambda_{f}|=1$ (No direct \CP violation).

\section{Radiative penguin decays}

$b \to s \gamma$ 
decays like $\B \to \Kstar \gamma$ are of particular interest because 
flavor changing neutral currents (FCNC) are forbidden in the standard model.
Effective FCNC are induced by one-loop penguins in which the top quark dominates.
Again, unknown heavy objects like charged Higgses or SUSY partners can
contribute in the loop, and the interference with the SM diagram can
induce \CP violating charge asymmetries as large as 20~\%, while the SM predicts
$A_{CP} < 1~\%$ \cite{ref:kagan}.

\begin{figure}[!htb]
\begin{center}
\includegraphics[width=0.6\linewidth]{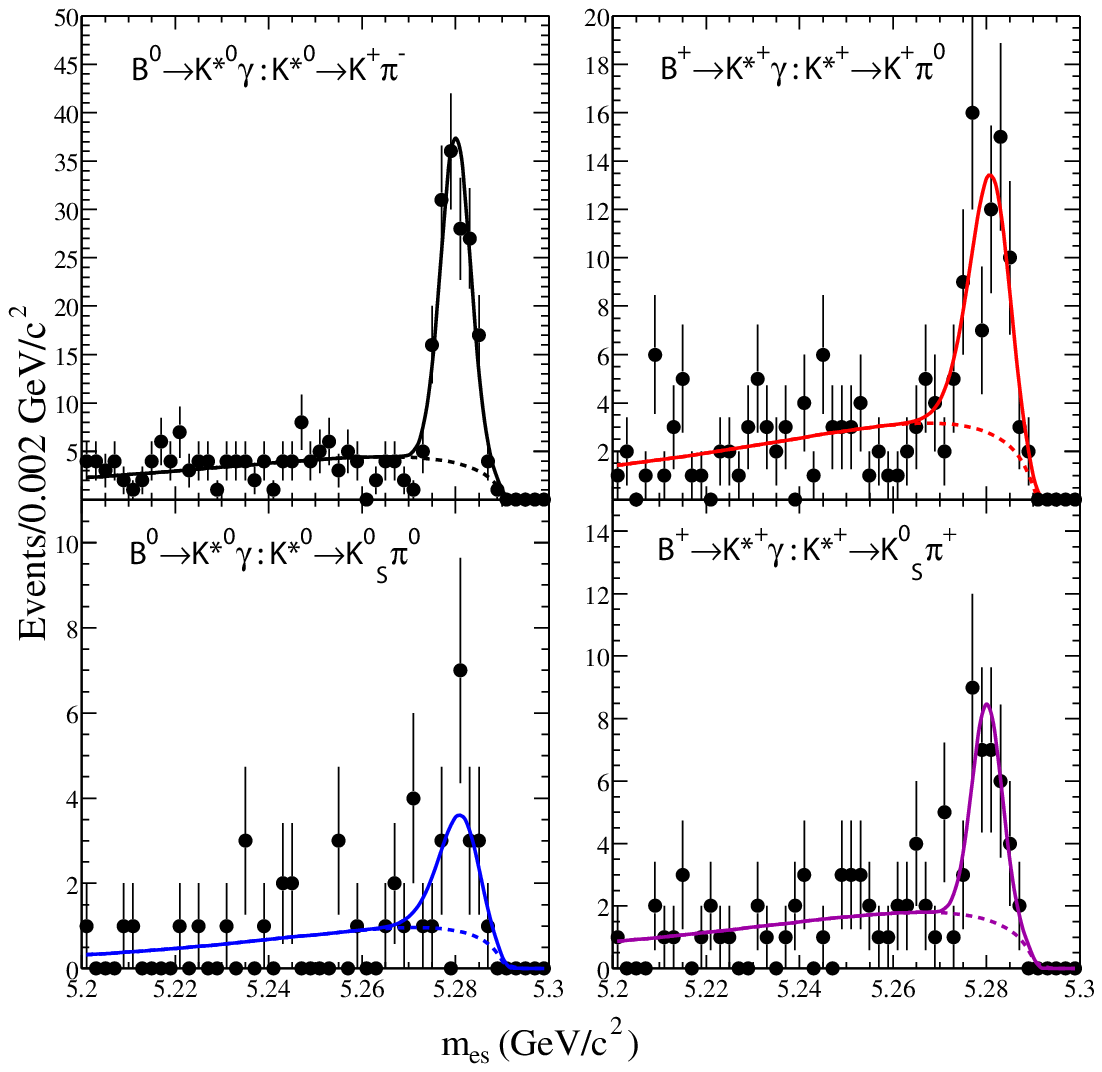}
\caption{ \mes distributions of the four $\Kstar \gamma$ channels (20.7 \invfb).
\label{fig:KstarGamma}}
\end{center}
\end{figure}

Monochromatic photons are selected ($2.30 < E_\gamma^* < 2.85\ \gev$).
As already noticed before, a large background from continuum events is present in this two body decay mode. 
The direction of the photon and the thrust of the rest of the event,
in the \FourS rest frame,
are uncorrelated for $\B \to \Kstar \gamma$ events, and strongly
correlated for background, and a cut on their respective angle
is the most powerful means of rejection available here.
The branching fractions are measured with an unbinned maximum
likelihood fit to the \mes distribution (fig. \ref{fig:KstarGamma},
table \ref{table:kstargamma}).

\begin{table}
\caption{Number of events, significance, and branching fraction of $\B \to \Kstar \gamma$ decays (20.7 \invfb).
\label{table:kstargamma}}
\begin{center}
\begin{tabular}{lcccc}
\hline \hline
 Mode & ${\cal B}(B\to K^*\gamma)/10^{-5}$ \\ \hline
\Kp\pim &  $4.39\pm 0.41\pm 0.27 $\\
\KS\piz &  $4.10\pm 1.71\pm 0.42 $\\
\KS\pip &  $3.12\pm 0.76\pm 0.21 $\\
\Kp\piz &  $5.52\pm 1.07\pm 0.33 $\\
\hline \hline
\end{tabular}
\end{center}
\end{table}
%

Charge asymmetry is measured for self-tagging modes (i.e. not
$\gamma\KS\piz$) and is found compatible with zero,
 $ A_{CP}= -0.035 \pm 0.076 \pm 0.012$: this measurement already 
constrains possible physics beyond the SM.

\section{Gluonic penguins}

The $\B\to\phi K^{(*)}$ decays present several interesting aspects:
the decay is dominated by a gluonic penguin contribution, making this mode
a smoking gun for the observation of this processes;
the \CP final eigenstate and the real decay amplitude
make the measurement of \stwob possible, in a system that has a different 
sensitivity to physics beyond the SM compared to $b \to c \bar{c} s$ decays 
\cite{GrossmannWorah}.

Branching fractions are measured (table \ref{table:phi}) with an
unbinned maximum likelihood fit to \mes, $\Delta E$, ${\cal F}$,
$m(\Kp\Km)$ (and $K\pi$ mass and $K^*$ helicity angle for $\phi K^*$
modes).
\begin{table}
\caption{$\B\to\phi K^{(*)}$ 
fitted number of signal event $n_{\hbox{sig}}$, statistical significance $S$,
and measured branching ratio ${\cal B}$. The subscripts in the $\phi
K^{*+}$ modes refer to the kaon daughter of the $\phi K^{*+}$ (20.7 fb$^{-1}$).
\label{table:phi}}
\begin{center}
\begin{tabular}{lcccc}
\hline \hline
Mode            & $n_{\hbox{sig}}$ & $S$ & ${\cal B}(10^{-6})$ \cr
\hline                          
$\phi K^+$      & $31.4^{+6.7}_{-5.9}$ & 10.5 &  $7.7^{+1.6}_{-1.4}\pm 0.8$ \cr
 $\phi K^0$     & $10.8^{+4.1}_{-3.3}$  & 6.4 &  $8.1^{+3.1}_{-2.5}\pm 0.8$ \cr
 $\phi K^{*+}$  & --  & 4.5 &  $9.7^{+4.2}_{-3.4}\pm 1.7$ \cr
\hline                          
~~$\phi K^{*+}_{K^+}$ & $7.1^{+4.3}_{-3.4}$ & 2.7 & $12.8^{+7.7}_{-6.1}\pm 3.2$ \cr
~~$\phi K^{*+}_{K^0}$ & $4.4^{+2.7}_{-2.0}$ & 3.6 & $ 8.0^{+5.0}_{-3.7}\pm 1.3$ \cr
\hline                          
$\phi K^{*0}$     & $20.8^{+5.9}_{-5.1}$ & 7.5 & $8.7^{+2.5}_{-2.1}\pm 1.1$ \cr
\hline
\hline
\end{tabular}
\end{center}
\end{table}
The decays $\Bp \to \phi K^{*+}$ and $\Bz \to \phi K^0$ are first
observations\cite{ref:phiKstar}.

\section{Summary}
\label{sec:Summary}

We have presented a sample of recent results of the \babar\ experiment.
\CP violation is observed in the \B sector with a significance of 
4.1 standard deviations.
The observation of several rare \B decays is reported, several of which allowed
a search for direct \CP violation in charge asymmetries.

The measurements presented here are statistically limited. 
The increase in integrated luminosity that is expected 
over the next few years -- 500 \invfb in year 2005 -- will allow a
significant improvement in the precision of these results.

\section{Acknowledgments}
\label{sec:Acknowledgments}

We are grateful for the 
extraordinary contributions of our \pep2\ colleagues in
achieving the excellent luminosity and machine conditions
that have made this work possible.
The collaborating institutions wish to thank 
SLAC for its support and the kind hospitality extended to them. 
This work is supported by the
US Department of Energy
and National Science Foundation, the
Natural Sciences and Engineering Research Council (Canada),
Institute of High Energy Physics (China), the
Commissariat \`a l'Energie Atomique and
Institut National de Physique Nucl\'eaire et de Physique des Particules
(France), the
Bundesministerium f\"ur Bildung und Forschung
(Germany), the
Istituto Nazionale di Fisica Nucleare (Italy),
the Research Council of Norway, the
Ministry of Science and Technology of the Russian Federation, and the
Particle Physics and Astronomy Research Council (United Kingdom). 
Individuals have received support from the Swiss 
National Science Foundation, the A. P. Sloan Foundation, 
the Research Corporation,
and the Alexander von Humboldt Foundation.



\begin{thebibliography}{99}

\bibitem{ref:babar}
\babar\ Collaboration, B.\ Aubert {\em et al.}, ``The \babar\ Detector, ''
\hepex{0105044} (2001), submitted to \nimBaseB

\bibitem{ref:wolf}
L.~Wolfenstein, 
\jprl{51}(1983)1945.

\bibitem{ref:prlfeb2001}
B. Aubert {\em et al.} \babar\ Collaboration, 
\jprl{86} 2515 (2001).

\bibitem{ref:prlsept2001}
B. Aubert {\em et al.} \babar\ Collaboration, 
\jprl{87}:091801 (2001).

\bibitem{ref:andreas}
A. H\"ocker {\em et al.}, \epjc{21} 225 (2001).

\bibitem{ref:charmlessAsym}
B. Aubert {\em et al.}, \babar\ Collaboration, 
\jprl{87}:151802 (2001).

\bibitem{ref:stwoa:th}
M. Gronau and D. London, \jprl{65}, 3381 (1990); \\
N.G. Deshpande and X.G. He, \jprl{74}, 26 (1995); \\
S. Gardner, \jprd{59}, 077502 (1999); \\
Y. Grossman and H.R. Quinn, \jprd{58}, 017504 (1998); \\
J. Charles, \jprd{59}, 054007 (1999); \\
M. Gronau, D. London, N. Sinha, and R. Sinha, \plb{514}315  (2001).

\bibitem{ref:stwoa}
Study of CP Violating Asymmetries in $\B \to \pip\pim, \Kp\pim$ Decays, 
By BABAR Collaboration (B. Aubert et al.). SLAC-PUB-8929, BABAR-CONF-01-05.
Contributed to 20th International Symposium on Lepton and Photon Interactions at High Energies 
(Lepton Photon 01), Rome, Italy, 23-28 Jul 2001. 
hep-ex/0107074.

\bibitem{ref:kagan}
A. Kagan and M. Neubert, \jprd{58} 094012 (1998).

\bibitem{ref:phiKstar}
B. Aubert {\em et al.}, \babar\ Collaboration, 
\jprl{87}:151801 (2001).


\bibitem{GrossmannWorah} Y. Grossman and M. P. Worah, 
\plb{395} 241 (1997).
\end{thebibliography}
\end{document}